\documentclass{rspublic}
\usepackage{graphicx}
\usepackage[numbers,sort&compress]{natbib}
\usepackage{amssymb,amsmath}

\usepackage[finalnew]{trackchanges}
\addeditor{RSP}

\begin{document}
\title[A new modelling framework for statistical cumulus dynamics]{A new modelling framework for statistical cumulus dynamics}
\author[R.\ S.\ Plant]{Robert S.\ Plant}
\affiliation{Department of Meteorology, University of Reading, PO Box 243, Reading, Berkshire RG6 2BB, UK.}
\label{firstpage}
\maketitle

\begin{abstract}{statistical cumulus dynamics, convective plumes, macroscopic limit, cumulus parameterization} 
We propose a new modelling framework suitable for the description of atmospheric convective systems as a collection of distinct plumes. The literature contains many examples of models for collections of plumes in which strong simplifying assumptions are made, \change{convective quasi-equilibrium and the macroscopic limit}{a diagnostic dependence of convection on the large-scale environment and the limit of many plumes} often being imposed from the outset. Some recent studies have sought to remove one or the other of those assumptions. The proposed framework removes both, and is explicitly time-dependent and stochastic in its basic character. The statistical dynamics of the plume collection are defined through simple probabilistic rules applied at the level of individual plumes, and van Kampen's system size expansion is then used to construct the macroscopic limit of the microscopic model. Through suitable choices of the microscopic rules, the model is shown to encompass previous studies in the appropriate limits, and to allow their natural extensions beyond those limits. \remove{Since the basic picture underlies most current convective parameterizations in general circulation and numerical weather prediction models, the framework provides a firm theoretical basis for investigations into improved parameterization approaches, and indeed for investigations of convective systems more generally.}  
\end{abstract}

\section{Introduction}\label{intro}

Any general-circulation model (GCM) of the Earth's atmosphere will contain a number of parameterizations of important processes that cannot be represented explicitly with the given discretization. These will include unresolved dynamics, such as boundary layer turbulence, and also physical processes, such as clouds and radiation. Stensrud \cite{stensrudbook} provides an overview of the methods currently in use in GCMs and weather forecast models. The problem of parameterization may ultimately be considered as that of developing a statistical description for the subgrid-scale processes, expressed in terms of their dependence on the known resolved-scale state. Conceptually at least, the procedure is analogous to that taken for deriving macroscopic thermodynamics from a microscopic description by means of statistical mechanics \cite{L+L}. For this reason, there is a lot to learn about parameterization from statistical mechanics, as well as statistical physics more generally, in order to develop subgrid-scale parameterizations in as robust a manner as possible.

The purpose of the present article is to contribute to the development of such a statistical approach for the parameterization of deep, precipitating convection. Deep convection is a crucial aspect of the tropical climate and deficiencies in its parameterization are implicated in some notoriously stubborn issues in climate modelling: some examples are equatorial waves \cite{lin+06}, the Madden-Julian oscillation \cite{lin08}, the spatial distribution of tropical rainfall \cite{ITCZ_aqua} and the diurnal cycle \cite{guichard}. Almost all current parameterizations \add{for GCMs} are based on an idealization of atmospheric convective systems as a collection of convective ``plumes'' \cite{AS74}. Each plume is embedded within and interacts with a horizontally--homogeneous medium called the environment. Thus the problem becomes how to construct a statistical cumulus dynamics (SCD) for the collection of plumes. 

Some rather brutal simplifying assumptions are made in translating even this idealized \change{system}{picture} into practical parameterizations \cite{arakawareview,stensrudbook,bulkreview}. We do not seek to review all the issues here, but rather focus upon two of the usual simplifications: first, the convection is assumed to be in equilibrium with a slowly-varying large-scale forcing; and, second, there are assumed to be very many plumes within a grid box of the parent GCM, such that an ensemble average is sufficient to represent the convective state. 
Some non-equilibrium models have been proposed in the literature which may describe aspects of the time evolution of convection \cite{panrandall,popdyn,p=1}\change{ and there have also been explorations of stochastic effects, either based on the intrinsic variability at equilibrium arising from finite cloud number \protect\cite{craigcohen,cohencraig,plantcraig}, or else more heuristically as attempts to account for generic parameterization uncertainty \protect\cite{varyparam,multnoise,generalstoch}.}{There have also been explorations of stochastic effects related to uncertainties in the triggering process \protect\cite{mk02,kmk03}, instrinsic flucuations at equilibrium arising from finite cloud number \protect\cite{craigcohen,cohencraig,plantcraig} and transitions in cloud morphology \protect\cite{kbm10}. Other treatments of stochastic effects have been introuduced for more heuristic reasons, as attempts to account for generic parameterization uncertainty \protect\cite{varyparam,multnoise,generalstoch}.}

Here, we propose a simple modelling framework suitable for the study of the statistical dynamics of cumulus clouds, which does not assume equilibrium and which treats stochastic effects due to finite cloud number. The framework has been sucessfully used for chemical and biological applications \cite{mckane04,pain,mckanebiochem}, but to the best of our knowledge, has not previously been exploited in atmospheric science. We will show that, in the appropriate limits, it agrees with \change{the just-cited}{previous} studies of both stochastic and time-varying aspects of convection. Moreover the model could easily be extended to permit future investigations of SCD with other assumptions removed: for example, a spatially-explicit form could be used to study structures arising from interactions of clouds with their local environment \cite{randallhuffman}.

The article is organized as follows. In Sec.~\ref{picture} we discuss the idealization of convective systems used as the basis of many current parameterizations, and \add{some} recent attempts to incorporate stochastic and time-varying aspects. The proposed modelling framework will be introduced in Sec.~\ref{ILM} and its relationship to the methods of Sec.~\ref{picture} will be analysed in some detail. Some examples of numerical results from the introduced models are presented in Sec.~\ref{numerical} and conclusions can be found in Sec.~\ref{concl}.

\section{Idealized picture for convection}\label{picture}

\subsection{The collection of plumes}\label{collection}

Following the usual idealization of convection in parameterization schemes \add{\protect\cite{arakawareview,gregory,tiedtke,kain90,bulkreview,AS74,plantcraig,popdyn}}, we consider a system of distinct cumulus clouds. Each cloud is described as a ``plume'' which is characterized by its mass flux, $M_i(z)$, defined by
\begin{equation}
M_i(z)=\rho \sigma_i w_i
\end{equation}
where the subscript labels the plume, $\rho$ is the density, $\sigma_i$ the fractional area occupied and $w_i$ the in-cloud vertical velocity. The mass flux is an important variable because it is assumed to dominate the sub-grid scale fluxes \cite{yano+04}. Denoting by $\chi$ some intensive variable of interest, its sub-grid flux due to convection is approximated by
\begin{equation}
\rho\overline{\chi'w'} \approx \sum_i M_i (\chi_i-\chi_\mathrm{env})
\end{equation}
the sum extending over all plumes present, the subscript env denoting the environmental value and the overbar and prime denoting respectively a horizontal average and a departure from that average. The approximation does require the fractional area occupied by cumulus clouds to be small, but typically this is not much larger than a few percent. \add{However, it should be noted that in some recent numerical weather prediction models, where the grid size approaches that of a convective element, then such an approximation becomes problematic \protect\cite[e.g.][]{gerard}.}

In order to compute $M_i(z)$ and $\chi_i(z)$ a description of the vertical structure \add{of each} plume is required, and various models have appeared in the literature \cite{AS74,kain90,emanuel_mono}. As the plume ascends, the in-plume and environmental air may interact, with mixing of some environmental air into the plume and of some in-plume air into the environment. There are long-standing debates about these interactions \cite{raymond+blyth}, which we do not revisit. Instead we will simply assume that some suitable plume model is available. Thus our concern will not be with the vertical structure of the plumes, but \change{with the question of}{rather with the magnitude of the convection: i.e.,} how many plumes are to be found within a given area for a given large-scale meteorological forcing? \remove{This is known as the closure problem in convective parameterization \protect\cite{AS74,stensrudbook}.}


Atmospheric convection is assumed to be forced by \remove{some} large-scale destabilization processes, \change{most commonly}{such as} radiative or advective cooling \add{or low-level moistening}. When convection occurs, it tends to restore stability, so that given enough time and a steady enough forcing a state of equilibrium may be achieved in which the forcing and convective tendencies are in balance. Convective quasi-equilibrium is the notion that the atmosphere is maintained close to such an equilibrium state \cite{AS74}, and for systems in quasi-equilibrium then the equilibrium level of convective activity can be imposed in \add{order to set a magnitude for the convection and so close} a parameterization \remove{as the closure condition}.  

\subsection{Concerning bulk models}

In actual parameterizations a common further simplification is the reduction of the system from a  collection of individual plumes to a single bulk plume \cite{gregory,tiedtke,kain90}. Real cumulus clouds have a wide variety of properties, and may, for instance, extend to different heights in the atmosphere. However, the equation sets typically used to describe a single plume are almost linear. A sum over plumes therefore recovers essentially the same form of equations as for the single plume \cite{yanai}, with the in-plume values $\chi_i$ of intensive variables replaced by their mass-flux weighted, or ``bulk'' values. This simplification does have some penalties \cite{bulkreview}, most notably the fact that very simple treatments of cloud microphysics and radiative effects are required for consistency. 

In the model proposed here, a single type of convective plume is considered, and accordingly we drop all plume subscripts in the following. However, it is important to recognize that this does not imply a bulk assumption. Rather it is done for simplicity and economy of presentation, in order that the main ideas should not be obscured. The extension of the model to multiple plume types is entirely straightforward. 

\subsection{Concerning finite cloud number}\label{stoch}

Assuming for the moment that an equilibrium state has been reached, we consider now the possibility of fluctuations about that state. Parameterizations usually neglect any such fluctuations, thereby implicitly assuming the number of clouds in a grid-box to be large. However, a simple scaling shows that fluctuations due to finite cloud number are likely to be far from negligible. Cloud-resolving model (CRM) studies give values for the mass flux at cloud base of $\sim 10^7$\,kg\,s$^{-1}$ \cite{cohencraig,davies} for a single moist convective plume\footnote{The cited studies used cloud definitions that will have encompassed shallow, non-precipitating clouds, and so are likely to give an underestimate for the deep, precipitating clouds which are the focus here. That point reinforces the argument given in the main text.}. The mass flux per unit area that is required to balance typical rates of forcing in the tropics can be estimated to be $\sim 10^{-2}$\,kg\,m$^{-2}$\,s$^{-1}$ \cite{p=1,randall+pan_mono}, so that for a GCM grid-box of size ($50$--$100$\,km)$^2$ only a few clouds can be expected to be present. \change{Similar}{Clearly the number density could be somewhat larger in places where clouds cluster together. However, rather similar} conclusions can be reached from coarse-graining calculations with CRM simulations \add{which do include convective organization} \cite{xu+,shuttspalmer07} or simply from visual inspection of satellite imagery. Convective instability is released in discrete events, the number of which is not sufficient on the scale of a GCM grid-box to produce a steady response to a steady forcing.

Fluctuations about a state of equilibrium can be described if it is known how the mass flux is partitioned amongst individual clouds \add{and how the clouds are spatially distributed}. \change{Cohen and Craig \protect\cite{craigcohen} showed}{Craig and Cohen \protect\cite{craigcohen} argued} that the partitioning can be determined from \add{the relevant} equilibrium statistical mechanics, by \change{treating the convective clouds analogously to a variable number of gas molecules in an ideal-gas system. The}{considering a variable number of non-interacting convective clouds subject to externally-imposed constraints. Their} predicted pdf for the total mass flux \add{within a finite area} is a convolution of a Boltzmann distribution for the mass flux per cloud and a Poisson distribution for the number of clouds present. These predictions have proved to be remarkably robust in CRM data \cite{cohencraig,shuttspalmer07,davies,davoudi}. The theory has been translated into a parameterization scheme \cite{plantcraig}, with the fluctuations about equilibrium having practical implications for the behaviour of the GCM \cite{ballplant}.  

Since this article considers a single plume type only, each cloud is assumed to have the same mass flux, and the \change{Cohen and Craig}{Craig and Cohen} theory reduces to the prediction of Poisson fluctuations at equilibrium.

\subsection{Concerning time-dependence and equilibrium}\label{prog}

For relatively rapidly varying forcings, the equilibrium assumption may break down, so that it becomes necessary to consider the time-dependence of convective  mass flux \cite{panrandall,davies}. However, even for steady forcing the evolution of a convective ensemble is a question of real interest. It is certainly not obvious a priori that a unique equilibrium state must be reached: the stationary state may be unstable, or multiple equilibria may be possible. 

The basic equations that have been used to consider the evolution of a convective ensemble were originally introduced by Arakawa and Schubert \cite{AS74}, and describe the energy cycle of the ensemble. Denoting by $A$ the vertical integral of in-plume buoyancy and by $K$ the vertically-integrated \add{convective} kinetic energy, these equations are
\begin{equation}
\frac{dA}{dt}=F - \gamma M_B \label{A_eqn}
\end{equation}
\begin{equation}
\frac{dK}{dt}=AM_B-\frac{K}{\tau_D} \label{K_eqn}
\end{equation}
$A$ is a measure of potential energy known as the cloud work function. Eq.~\ref{A_eqn} shows it to be generated through the action of large-scale forcing $F$ and removed through the presence of convection. The quantity $\gamma$ gives the removal rate per unit of cloud-base mass flux, $M_B \equiv M(z=z_\mathrm{base})$. Both $F$ and $\gamma$ are calculable for any given plume model, and from such calculations it is known that the removal of instability is usually dominated by warming due to compensating environmental subsidence \cite{AS74,yanai}. The cloud work-function must be positive in order for convective kinetic energy to be generated, as shown by the first term on the right-hand side of Eq.~\ref{K_eqn}. Kinetic energy is assumed to be removed through a dissipation term, for which the value of the dissipation timescale $\tau_D$ is somewhat controversial. $\tau_D$ has been variously estimated to be in the range $10^3$ to $10^6$\,s.

Eqs.~\ref{A_eqn} and \ref{K_eqn} could be integrated if there were some functional relationship between $K$ and $M_B$. One relationship which has been postulated \cite{randall+pan_mono,panrandall} is that
\begin{equation}
K = \alpha M_B^2 \label{p=2eqn}
\end{equation}
with $\alpha$ treated as a constant. This would not appear implausible at first sight, given that \change{$K \sim \sigma w^2$}{$K \sim \rho \sigma w^2$} and \change{$M_B \sim \sigma w$}{$M_B \sim \rho \sigma w$}. For a single plume type with steady forcing, the postulate gives rise to a damped harmonic oscillator that approaches equilibrium after a few $\tau_D$ \cite{randall+pan_mono}. Recently, Yano and Plant \cite{p=1} have argued that Eq.~\ref{p=2eqn} is inconsistent with CRM results and theoretical scalings \add{for the dependencies of the equilibrium state \protect\cite[e.g.][]{emanuel+bister,shutts+gray}}, which better support their postulate of
\begin{equation}
K=\beta M_B \label{p=1eqn}
\end{equation}
with $\beta$ treated as a constant. \add{This is effectively an assumption that the response of deep convection to variations in the large-scale forcing occurs mainly through variations in fractional area rather than through typical in-cloud velocities.} For a single plume type with steady forcing, \change{this}{the} postulate gives rise to a periodic orbit, with cycles of convective recharge and discharge. Note, however, that the orbit is structurally unstable, such that small changes to the model typically produce a slow spiral in $AM_B$ space towards the equilibrium state. 

A third prognostic system based on Eqs.~\ref{A_eqn} and \ref{K_eqn} has also been proposed \cite{popdyn}, and is analogous to the Lotka-Volterra equations of population dynamics, with the clouds competing to consume convective instability. 
\begin{equation}
A \frac{dM_B}{dt}= F M_B - \gamma M_B^2 \label{popdyn}
\end{equation}
The form of this system is actually rather insensitive to the postulated relationship between $K$ and $M_B$ but it does require an additional assumption that Eq.~\ref{K_eqn} approaches equilibrium much more rapidly than does Eq.~\ref{A_eqn}. Further discussion of these points is given by Plant and Yano \cite{popdyncomment}.

The population dynamics system \cite{popdyn} provides a good illustration of how the consideration of prognostic systems may prove instructive for GCM parameterizations, even those for which convective quasi-equilibrium is imposed. Shallow, non-precipitating cumulus clouds are typically treated somewhat differently from deep, precipitating clouds in a GCM, possibly through parameter differences within a parameterization, or possibly through the use of different parameterization schemes entirely. Whether a given GCM considers convection to be shallow or deep rests on physically-motivated but undeniably somewhat ad hoc criteria. However, should convective ensembles prove to be well described by Lotka-Volterra equations it then follows that, for two convective types, a globally--stable equilibrium state with co-existing shallow and deep clouds exists if and only if the known coefficients $A$, $F$ and $\gamma$ statisfy certain inequalities \cite{takeuchi}. Otherwise, one of the convective types must be driven to extinction. Thus, in an equilibrium-based parameterization, clear criteria would dictate which scheme(s) to apply.

\section{Individual-Level Model}\label{ILM}

The stochastic model discussed in Sec.~\ref{picture}\ref{stoch} assumed convective equilibrium, whilst the three prognostic systems discussed in Sec.~\ref{picture}\ref{prog} all assumed an infinite number of clouds to be present (this is necessary in order for $M_B$ to be continuous and $dM_B/dt$ well defined). We now propose a modelling framework for statistical cumulus dynamics which is both stochastic and prognostic. The basic system is formulated in terms of an extremely simple set of probabilistic rules at the level of individual clouds, which are born, interact with their environment through changes to its cloud work-function, and die. According to our choice of these rules, we can produce models that are the microscopic analogues to any of the prognostic systems above.

\change{Indeed, we will show this by using}{Stochastic birth-death processes have previously been used to describe deep convection \protect\cite{mk02,kmk03,kbm10}, with some encouraging results when coupled to idealized models of large-scale tropical dynamics \protect\cite{mfk08,fmk11}. Here the aim and the context is somewhat different. We will assume the large-scale tendencies to be externally prescribed, in the tradition of idealized CRM and single-column model experiments that have long been used to develop parameterizations for operational weather forecast and climate models. We then seek to make direct links between the microscopic model and the above prognostic systems in a suitable limit, establishing in particular which microscopic processes are required, are admissible, or are forbidden in order to make contact with each of the prognostic systems. As discussed above, stochastic and prognostic aspects of convective parameterization are attracting increasing attention and may be starting to produce promising results \protect\cite{panrandall,varyparam,multnoise,popdyn}. The objective here is to show how those aspects might be combined in a natural way that is consistent with existing studies.

Specifically, we will use} van Kampen's expansion \cite{vankampen} to recover the ordinary differential equations of Sec.~\ref{picture}\ref{prog} as the macroscopic, large system-size limits of individual-level models. The leading correction for a non-infinite system is a Fokker-Planck equation describing the fluctuations in $N$ and $A$. A detailed analysis of those fluctuations is beyond the scope of the present article, but clearly offers promise for extending the validity of stochastic convective parameterizations and possibly also for developing theoretical interpretations of observational data \cite{peters1f,yano1f}.

\subsection{Definition of modelling framework}\label{micro_define}

The microscopic model is described through the probability distribution function $P(N,A,\tau)$ for the number of clouds $N$ and the cloud work function $A$ at time $\tau$. The domain of interest contains $\Omega$ elements, each element being defined as the minimum area necessary to support a single cumulus cloud. It is not necessary to specify a numerical value for the area, but we could consider $\sim$($1$--$5$\,km)$^2$ to be reasonable. \add{The area elements are not labelled and all elements are considered to have an equal chance of interacting with each other. In other words, no account is taken of whether interacting elements are nearest neighbours or well separated. The model could be generalized to include spatial dependence, with the van Kampen expansion used to derive corresponding macroscopic partial differential equations. However, since such spatial aspects are not treated in the comparison studies of Secs.~\protect\ref{picture}\protect\ref{stoch},\protect\ref{picture}\protect\ref{prog} they will similarly be regarded as out of scope here.} 

\change{$P$}{The model} evolves according to state transition probabilities that represent births and deaths, and environmental destabilization and stabilization. This evolution is governed by a master equation,
\begin{equation}
\frac{\partial P(N,A,\tau)}{\partial \tau}=\int dA' \sum T(N,A|N',A')P(N',A',\tau)- T(N',A'|N,A)P(N,A,\tau) \label{master}
\end{equation}
The transition matrix elements $T(f|i)$ denote the probability per unit time of making a transition from an initial state $i$ to a final state $f$. The master equation is therefore simply a statement of balance for the probability of state $(N,A)$: the first term on the right-hand side of Eq.~\ref{master} represents a gain in probability due to transitions to the state of interest, while the second term represents a loss of probability due to transitions from the state of interest. 

We now define the processes that can lead to transitions in the state of the system. In considering a possible transition, we may choose to look at either one or two \change{sites on the lattice}{elements}, with probabilities $1-\mu$ and $\mu$ respectively. Let us suppose that we look at one \change{site}{element}. The total number of \change{sites on the grid}{elements} is designated by $\Omega$, \change{and the grid contains $N$ sites}{of which $N$ elements are} occupied by clouds and $E=\Omega-N$ \change{empty sites}{elements are empty}. Simple combinatorics dictates that the chance of the single \change{site}{element} being currently occupied is $N/\Omega$ while the chance that it is currently empty is $E/\Omega$.

Let us now suppose that the single \change{site}{element} chosen is currently empty. The \change{site}{element} may become occupied through the formation of a cloud and we denote the probability of this happening as $a$. We might anticipate a dependence of $a$ on the cloud work function $A$, with cloud formation being more likely for larger $A$, but let us reserve judgement for the moment on any such dependence. On the other hand, if no cloud is formed then the maintenance of an empty \change{site}{element} will contribute to destabilization of the atmosphere by large-scale forcing, which may be represented through an increment of $s$ in the cloud work function. From these considerations, and combining the relevant probabilistic factors, we can now write down elements of the transition matrix as follows
\begin{equation}
T(N+1,A|N,A')=a(1-\mu) \frac{E}{\Omega} \delta (A-A') + \ldots \label{spon_birth}
\end{equation}
\begin{equation}
T(N,A|N,A')=(1-a)(1-\mu) \frac{E}{\Omega} \delta (A-A'-s) + \ldots \label{destab}
\end{equation}
the dots indicating that there are additional contributions.

In Table~\ref{tab:processes} we specify all processes \add{that will be} considered \add{as physically plausible} in the individual-level model. The processes of cloud formation and large-scale environmental forcing that were just discussed are listed in the first two lines. We refer to this particular process of cloud formation as spontaneous birth in order to distinguish it from other possible formation processes. \add{All of the mathematical expressions appearing in the Table are simply composed of the products of appropriate probabilistic factors, constructed analogously to those appearing in Eqs.~\protect\ref{spon_birth} and~\protect\ref{destab}.} 

\change{If an occupied site is sampled then it might die (with probability $d$), but otherwise its maintenance will stabilize the atmosphere through a reduction of $r$ in the cloud work function. Should two sites be sampled then two unoccupied sites may give birth to a cloud (with probability $e$), two occupied sites may cause a cloud to be eliminated through competitive exclusion (with probability $c$) while one occupied and one unoccupied site  may induce the birth of a cloud (with probability $c$). The last of these processes could correspond physically to triggering at the edge of a cold pool produced by pre-existing convection \protect\cite{tompkins}. Should two  sampled be selected but the number of clouds not change then an appropriate change may be made to the cloud work function, as listed in Table~\ref{tab:processes}.}{Suppose that a single element is selected and is found to be occupied. The cloud might die (with probability $d$), but otherwise its continued existence will stabilize the atmosphere through a reduction of $r$ in the cloud work function. Supposing that two elements are sampled then they could be both unoccupied, both occupied, or else one is occupied and the other is not. If two unoccupied elements are chosen then we allow for the possible birth of a cloud (with probability $e$); if two occupied elements are chosen then we allow for the possible death of a cloud through competitive exclusion (with probability $c$); while if one occupied and one unoccupied element is chosen then we allow that the pre-existing cloud may induce the birth of a new cloud (with probability $b$). The last of these processes could correspond physically to triggering at the edge of a cold pool produced by pre-existing convection \protect\cite{tompkins}. Should two elements be sampled  but the number of clouds not change through one of the above processes then an appropriate change is made to the cloud work function. Analogously to the changes in cloud work function for the case of a single sampled site, the maintenance of each unoccupied element destabilizes the atmosphere through an increment $s$ whereas the maintenance of each occupied element stabilizes the atmosphere through a reduction $r$.}

\begin{table}
\caption{Processes considered in the individual-level model. The second column indicates the \change{site}{element}(s) chosen for consideration, with O and U denoting an occupied and an unoccupied \change{site}{element} respectively. All other notation is specified in the main text.}
\label{tab:processes}
\begin{tabular} {|l|l|l|l|}
\hline
Process name & \change{Site}{Element}(s) & Matrix element & Value \\
\hline
spontaneous birth & U & $N+1,A|N,A'$ & $a(1-\mu) E\Omega^{-1} \delta (A-A')$\\
destablization & U & $N,A|N,A'$ & $(1-a)(1-\mu) E\Omega^{-1} \delta (A-A'-s)$\\
death & O & $N-1,A|N,A'$ & $d(1-\mu) N\Omega^{-1} \delta (A-A')$\\
stabilization & O & $N,A|N,A'$ & $(1-d)(1-\mu) N\Omega^{-1} \delta (A-A'+r)$\\
induced birth & UO & $N+1,A|N,A'$ & $2b\mu EN \Omega^{-1}(\Omega-1)^{-1} \delta (A-A')$\\
modification & UO & $N,A|N,A'$ & $2(1-b)\mu EN \Omega^{-1}(\Omega-1)^{-1} \times$\\ 
 & & & $\times \delta (A-A'-s+r)$\\
exclusion & OO & $N-1,A|N,A'$ & $c\mu N(N-1) \Omega^{-1}(\Omega-1)^{-1} \delta (A-A')$\\
strong stabilization & OO & $N,A|N,A'$ & $(1-c)\mu N(N-1) \Omega^{-1}(\Omega-1)^{-1} \times$\\
 & & & $\times \delta (A-A'+2r)$\\
birth & UU  & $N+1,A|N,A'$ & $e\mu E(E-1) \Omega^{-1} (\Omega-1)^{-1} \delta (A-A')$\\
strong destabilization & UU & $N,A|N,A'$ & $(1-e)\mu E(E-1) \Omega^{-1} (\Omega-1)^{-1} \times$\\
 & & & $\times \delta (A-A'-2s)$\\
\hline
\end{tabular}
\end{table}

Natural boundary conditions to impose upon the transition matrix elements are the requirements that $T(-1,A|0,A')=0$ and $T(\Omega+1,A|\Omega,A')=0$ so that by starting the system in a physical configuration with $0\le N\le \Omega$ it will not be able to reach an unphysical configuration. Those conditions are satisfied by the expressions presented in Table~\ref{tab:processes}. 

\subsection{System Size Expansion}

In order to make contact between the individual-level, probabilistic model and the deterministic ordinary differential equations of Sec.~\ref{picture}\ref{prog} we perform the system size expansion of van Kampen \cite{vankampen}. Detailed demonstrations of the method are available elsewhere \cite{vankampen,mckane04,mckanebiochem,pain} but as it may be unfamiliar to atmospheric scientists, an outline will be presented here. For illustrative purposes, we will focus our attention on the spontaneous birth and environmental destabilization processes of Eqs.~\ref{spon_birth} and \ref{destab}, but the manipulations for the other processes in Table~\ref{tab:processes} follow along very similar lines.
 
For a large-enough, horizontally-homogenous domain we would expect the cloud work function to be almost independent of the system size $\Omega$, albeit with some small fluctuations. \add{The central limit theorem suggests that such fluctuations would be} of order $1/\sqrt{\Omega}$. \add{Our simulations of the individual-based models presented here also support such a scaling.} The essence of the system-size expansion is \add{to assume this scaling and so} decompose the cloud work function into a macroscopic, size-independent, determinstic part $\varphi$ and a fluctuating, stochastic part $\lambda$. Thus, 
\change{
\protect\begin{equation}
A(t) = \varphi(t)+\Omega^{-1/2} \lambda(t) 
\nonumber
\protect\end{equation}
}{
\protect\begin{equation}
A(\tau) = \varphi(\tau)+\Omega^{-1/2} \lambda(\tau) \label{a_decompose}
\protect\end{equation}
}
Similar considerations apply to the number of clouds present, although we would expect this to scale with the system size in the macroscopic limit. Thus, the decomposition takes the form
\change{
\protect\begin{equation}
N(t)=\Omega \sigma(t)+\Omega^{1/2} \eta(t) 
\nonumber
\protect\end{equation}
}{
\protect\begin{equation}
N(\tau)=\Omega \sigma(\tau)+\Omega^{1/2} \eta(\tau) \label{n_decompose}
\protect\end{equation}
}
\remove{Note that the time $t$ appearing in these decompositions is a macroscopic timescale to be defined later.}

To apply the above decomposition to the master equation, we introduce in place of $P(N,A,\tau)$ a function \change{$\Pi (\eta,\lambda,t)$}{$\Pi (\eta,\lambda,\tau)$} which will describe the probabilities for the fluctuating variables. \change{The partial derivative of $P$ on the left-hand side of Eq.~\protect\ref{master} is to be taken at constant $N$ and $A$, and so may be transformed as follows}{Considering the left-hand side of Eq.~\protect\ref{master}, the chain rule immediately gives
\begin{equation}
\frac{\partial P}{\partial \tau}=\frac{\partial \Pi}{\partial \tau}+ \frac{\partial \Pi}{\partial \eta} \frac{d\eta}{d\tau}+\frac{\partial \Pi}{\partial \lambda} \frac{d\lambda}{d\tau} 
\end{equation}
and since the time derivatives of the fluctuating variables are to be taken with $N$ and $A$ held constant, Eqs.~\protect\ref{a_decompose} and~\protect\ref{n_decompose} can be used to relate them to the time derivatives of the macroscopic variables. The result is that}
\begin{equation}
\frac{\partial P}{\partial \tau}=\frac{\partial \Pi}{\partial \tau}-\Omega^{1/2} \frac{d\sigma}{d\tau} \frac{\partial \Pi}{\partial \eta}-\Omega^{1/2} \frac{d\varphi}{d\tau} \frac{\partial \Pi}{\partial \lambda} \label{transform_pdf}
\end{equation}

The state transitions in the master equation can be expressed in terms of ladder operators for changes in cloud number and cloud work function. For some arbitrary function $f(N,A)$ these are defined by
\begin{equation}
\Upsilon^\pm f(N,A) = f(N\pm1,A) 
\end{equation}
\begin{equation}
\Gamma^q f(N,A) = f(N,A+q) 
\end{equation} 
These operators can be expanded in powers of $\Omega$, reflecting the fact that a single transition in a large system will induce only a small change in the fluctuating variables. Specifically, 
\begin{equation}
\Upsilon^\pm=1 \pm \Omega^{-1/2} \frac{\partial}{\partial \eta} +\frac{1}{2} \Omega^{-1} \frac{\partial^2}{\partial \eta^2} \pm \ldots \label{expand_nladder}
\end{equation}
\begin{equation}
\Gamma^q=1 + \Omega^{-1/2} q \frac{\partial}{\partial \lambda} +\frac{1}{2} \Omega^{-1} q^2 \frac{\partial^2}{\partial \lambda^2} + \ldots \label{expand_aladder}
\end{equation}

\add{Table~\protect\ref{tab:processes} describes processes associated with just one or two elements of the microscopic model. The macroscopic model is assumed to be much larger and the intensive variables $\varphi$ and $\sigma$ describing it will evolve more slowly, in response to changes that have affected the full set of elements. It is therefore convenient to introduce a macroscopic time $t$ through a rescaling of the microscopic time, setting}
\remove{The final ingredient necessary to relate the microscopic and macroscopic descriptions of the system is to connect the two timescales by setting}
\change{
\protect\begin{equation}
t=\Omega \tau 
\nonumber
\protect\end{equation}
}{
\protect\begin{equation}
t=\Omega^{-1} \tau \label{trescale}
\protect\end{equation}
}
\change{Since the cloud work function is an intensive variable in the macroscopic limit the changes to it due to a single transition will be of order $\Omega^{-1}$. It is therefore convient to rescale the quantities which alter the cloud work function in the individual-level model, by writing}{Similarly the quantities describing a change to the cloud work function from one or two elements of the microscopic model are also rescaled,}  
\begin{equation}
\tilde{r}=r\Omega \quad ; \quad \tilde{s}=s\Omega \label{rs_rescale}
\end{equation}
so that $\tilde{r}$ and $\tilde{s}$ are quantities of order $\Omega^0$.

Let us now substitute Eqs.~\ref{a_decompose}-\ref{transform_pdf}, \ref{rs_rescale} and \ref{trescale} into the master equation, and also make use of the ladder operator expansions of Eq.~\ref{expand_nladder} and \ref{expand_aladder}. This leads to the following contributions for the example processes
\begin{gather}
\Omega^{-1} \left[ \frac{\partial \Pi}{\partial t}-\Omega^{1/2} \frac{d\sigma}{dt} \frac{\partial \Pi}{\partial \eta}-\Omega^{1/2} \frac{d\varphi}{dt} \frac{\partial \Pi}{\partial \lambda} \right] = \ldots  \nonumber \\
+ \left[ -\Omega^{-1/2} \frac{\partial}{\partial \eta}+\Omega^{-1}\frac{1}{2}\frac{\partial^2}{\partial \eta^2}+\ldots \right] a(1-\mu) \frac{1}{\Omega} (\Omega-\Omega\sigma-\Omega^{1/2}\eta) \Pi \nonumber \\
+ \left[ -\Omega^{-1/2} \tilde{s}\frac{\partial}{\partial \lambda}+\Omega^{-1}\frac{1}{2}\tilde{s}^2\frac{\partial^2}{\partial \lambda^2}+\ldots \right] (1-a)(1-\mu) \frac{1}{\Omega} (\Omega-\Omega\sigma-\Omega^{1/2}\eta) \Pi 
\label{substitute}
\end{gather}
Collecting together the terms at the leading order in $\Omega$ this gives 
\begin{equation}
-\frac{\partial \Pi}{\partial \eta} \frac{d\sigma}{dt} -\frac{\partial \Pi}{\partial \lambda} \frac{d\varphi}{dt} = 
- \frac{\partial}{\partial \eta} a(1-\mu) (1-\sigma) \Pi 
-\tilde{s}\frac{\partial}{\partial \lambda} (1-a)(1-\mu) (1-\sigma) \Pi +\ldots 
\label{LO}
\end{equation}
Consider now the action of the derivative operators on the right-hand side of the above equation. Recall from Sec.~\ref{ILM}\ref{micro_define} that we suggested that it may be appropriate for the transition probability $a$ to have some dependence on the cloud work function $A$. \change{In so far as $a$ depends on $\varphi$, the macroscopic part of $A$, such a dependence would not give rise to any contributions to Eq.~\protect\ref{LO}. In so far as $a$ depends on $\lambda$, the fluctuating part of $A$, these would lead to contributions from $\partial a/\partial \lambda$ but as such contributions are suppressed by a factor of $\Omega^{-1/2}$ they are not relevant at the leading-order level of Eq.~\protect\ref{LO}. It follows that the derivatives on the right-hand side of Eq.~\protect\ref{LO} may be considered to act on $\Pi$ only.}{From the chain rule, $\partial a/\partial \lambda = (\partial A/\partial \lambda)(da/dA)= \Omega^{-1/2} da/dA$, the decomposition of $A$ from Eq.~\protect\ref{a_decompose} having been used in the final equality. Thus, any dependence of $a$ on $A$ is irrelevant at the leading-order level of Eq.~\protect\ref{LO} and the derivatives on the right-hand side of that equation may be considered to act on $\Pi$ only.} 

In order to satisfy the leading-order equation it is sufficient that the macroscopic functions $\sigma$ and $\varphi$ should obey \remove{the} ordinary differential equations \add{that can be obtained by equating the respective coefficients of $\partial \Pi/\partial \eta$ and $\partial \Pi/\partial \lambda$ in Eq.~\protect\ref{LO}. These are:} 
\begin{equation}
\frac{d\sigma}{dt} = e\mu + a(1-\mu) + \sigma \left[2(b-e)\mu - (a+d)(1-\mu) \right] + \sigma^2 \mu (e-2b-c) \label{sigma_full}
\end{equation}
\begin{gather}
\frac{d\varphi}{dt} =  \tilde{s}\left[ 2(1-e)\mu + (1-a)(1-\mu) \right] \nonumber \\
 + \sigma \left[ 2(\tilde{s}-\tilde{r})(1-b)\mu -4\tilde{s}(1-e)\mu - \tilde{s}(1-a)(1-\mu) -\tilde{r}(1-d)(1-\mu)  \right] \nonumber \\
+2 \sigma^2 \mu \left[ \tilde{s}(1-e) - (\tilde{s}-\tilde{r})(1-b)  -\tilde{r}(1-c)  \right] \label{A_full}
\end{gather}
where $a, \ldots e, \mu, \tilde{r}$ and $\tilde{s}$ could be considered as functions of $\sigma$ and $\varphi$. Notice that we have here stated explicitly the contributions from all of the processes listed in Table~\ref{tab:processes}.

The terms at next-to-leading order in Eq.~\ref{substitute} contain powers of ${\cal O} (\Omega^{-1})$. Inspection of Eq.~\ref{substitute} shows that they will take the form of a Fokker-Planck equation for $\Pi$, which is easily derived but not stated here.


\subsection{Relation to time-dependent convection models}

We now discuss the connections from the macroscopic equations derived in the previous subsection with models that have been proposed for the prognostic description of atmospheric convection, as presented in Sec.~\ref{picture}\ref{prog}.

In order to establish such a connection, it is important to recall from Sec.~\ref{picture}\ref{collection} that the mass-flux approximation for the description of convective plumes approximates the fractional area occupied by clouds as being small. To make an appropriate comparison to the individual-level model, the equivalent approximation should also be made there. This corresponds to setting $E \approx \Omega$ in all of the transition matrix elements and results in macroscopic equations that are then reduced to the following.
\begin{equation}
\frac{d\sigma}{d\tau} = e\mu + a(1-\mu) + \sigma \left[ 2b\mu  - d(1-\mu) \right] - c\mu \sigma^2 \label{sigma_lim}
\end{equation}
\begin{gather}
\frac{d\varphi}{d\tau}=\tilde{s}\left[ 2(1-e)\mu + (1-a)(1-\mu) \right]
 + \sigma \left[ 2(\tilde{s}-\tilde{r})(1-b)\mu -\tilde{r}(1-d)(1-\mu)  \right] \nonumber \\
-2 \sigma^2 \mu \tilde{r}(1-c) \label{A_lim}
\end{gather}

\change{Now}{First} let us consider the population dynamics system of Wagner and Graf \cite{popdyn}. This system does not consider the evolution of the cloud work function so all processes in Table~\ref{tab:processes} involving changes to $A$ should be neglected. With only a single plume type being considered, the macroscopic fractional cloud number is proportional to cloud-base mass flux, and hence Eqs.~\ref{popdyn} and \ref{sigma_lim} may be compared directly. To reproduce the structure of Eq.~\ref{popdyn} from the individual-based model one simply includes the induced birth and competitive exclusion processes from Table~\ref{tab:processes}. Values for all parameters of the microscopic model could then be set immediately from the known coefficients of Eq.~\ref{popdyn}. This would constitute a minimal microscopic model fully consistent with the macroscopic population dynamics system. Indeed, exactly such a population model has been considered in other contexts, including spatial effects and interacting types \cite{mckane04}. Notice that inclusion of the death process from Table~\ref{tab:processes} would produce a more complicated microscopic model that would again be entirely consistent with macroscopic population dynamics, although the character of the fluctuations would be rather different. However, the birth processes from one or two empty \change{sites}{elements} may not be included.

We turn now to the prognostic convective system defined by Eqs.~\ref{A_eqn} and \ref{K_eqn} with the Pan and Randall postulate \cite{panrandall,randall+pan_mono} of Eq.~\ref{p=2eqn}. To obtain this system the quadratic terms in $\sigma$ appearing in Eqs.~\ref{sigma_lim} and \ref{A_lim} must be eliminated by neglecting the microscopic model processes involving two occupied \change{sites}{elements}. All other processes listed in Table~\ref{tab:processes} coud be retained if desired, but for a minimal microscopic model it is necessary to retain only those processes for which a single \change{site on the lattice}{element} is sampled. \add{Notice that the probability for spontaneous birth must be chosen to be proportional to the cloud work function.}

The final prognostic system of interest is that defined by Eqs.~\ref{A_eqn} and \ref{K_eqn} with the Yano and Plant postulate \cite{p=1} of Eq.~\ref{p=1eqn}. As for the previous system, microscopic processes involving two occupied \change{sites}{elements} are neglected. It is also necessary, however, to eliminate the constant term from Eq.~\ref{sigma_lim} by neglecting the process in Table~\ref{tab:processes} of spontaneous birth and also that of birth arising from two empty \change{sites}{elements}. Furthermore, to obtain the correct structure, \add{in this case} the probability $b$ for the induced birth process must be chosen to be proportional to the cloud work function, a choice which then necessitates the neglect of the modification process from Table~\ref{tab:processes}. These decisions could also be supplemented by the optional neglect of strong destabilization to arrive at the minimal microscopic model corresponding to the prognostic system.

Table~\ref{tab:summary} summarizes the various forms of the individual-level model that are required in order to produce the three macroscopic convective systems in the limit of large system size.

\begin{table}
\caption{A list of processes from the individual-level model and their roles in producing three macroscopic convective systems. Pop dyn refers to the population dynamics system of Wagner and Graf \cite{popdyn}; PR refers to the system of Pan and Randall \cite{panrandall,randall+pan_mono} with $K\sim M^2$; and, YP refers to the system of Yano and Plant \cite{p=1} with $K\sim M$.}
\label{tab:summary}
\begin{tabular} {|l|c|c|c|}
\hline
Process name & Pop dyn & PR & YP \\
\hline
spontaneous birth & neglect & include & neglect\\
destablization & neglect & include & include\\
death & optional & include & include\\
stabilization & neglect & include & include\\
induced birth & include & optional & include\\
modification & neglect & optional & neglect\\ 
exclusion & include & neglect & neglect\\
strong stabilization & neglect & neglect & neglect\\
birth & neglect  & optional & neglect\\
strong destablization & neglect & optional & optional\\
\hline
\end{tabular}
\end{table}

\section{Numerical Results}\label{numerical}

In this section we present some example results obtained from the three minimal individual-level models that in the large system-size limit are equivalent to the three prognostic models of convective systems described in Sec.~\ref{picture}\ref{prog}. 

For each case we choose consistent macroscopic parameters, in the sense that the systems have the same equilibrium state in the mass-flux limit of a vanishing fractional cloud area. The values taken are consistent with the ranges found in the cited literature: specifically we have set $F=10^{-2}$\,m$^2$\,s$^{-3}$, $\gamma=1$\,m$^4$\,kg$^{-1}$\,s$^{-2}$, $\tau_D=10^3$\,s, $\beta=5\times10^4$\,m$^2$\,s$^{-1}$ and $\alpha=5\times10^6$\,m$^4$\,kg$^{-1}$. The proportionality factor connecting $M_B$ and $\sigma$ we have set to $0.1$\,kg\,m$^{-2}$\,s$^{-1}$. This particular choice is a somewhat small value that will perhaps overestimate the number of clouds required in order to produce the equilibrium level of mass flux. However, it is convenient in that it allows us easily to compare results from the individual-based models with those from the macroscopic convective systems, without having to take a very large system size or very many realizations of the probabilistic model. For these illustrations $\Omega=1000$, and 100 realizations have been simulated.

The above macroscopic parameters are sufficient to determine all of the relevant parameters for the equivalent minimal individual-based models of both the population dynamics system \cite{popdyn} and the Pan and Randall system \cite{panrandall,randall+pan_mono}. For the minimal equivalent to the Yano and Plant system \cite{p=1}, one microscopic parameter remains undetermined. We have chosen this to be $\mu$ and have assigned it the arbitrary value of $0.1$. It cannot be determined from macroscopic considerations alone, but should properly be set from investigation of the fluctuations in those convective systems that are well described by the model. Of course, the same remark holds for all three models in respect of whether and how any non-minimal processes should also be included in order to account more fully for convective fluctuations. 


\begin{figure}[htb]
\centering
\includegraphics[width=0.3\textwidth,height=0.25\textheight]{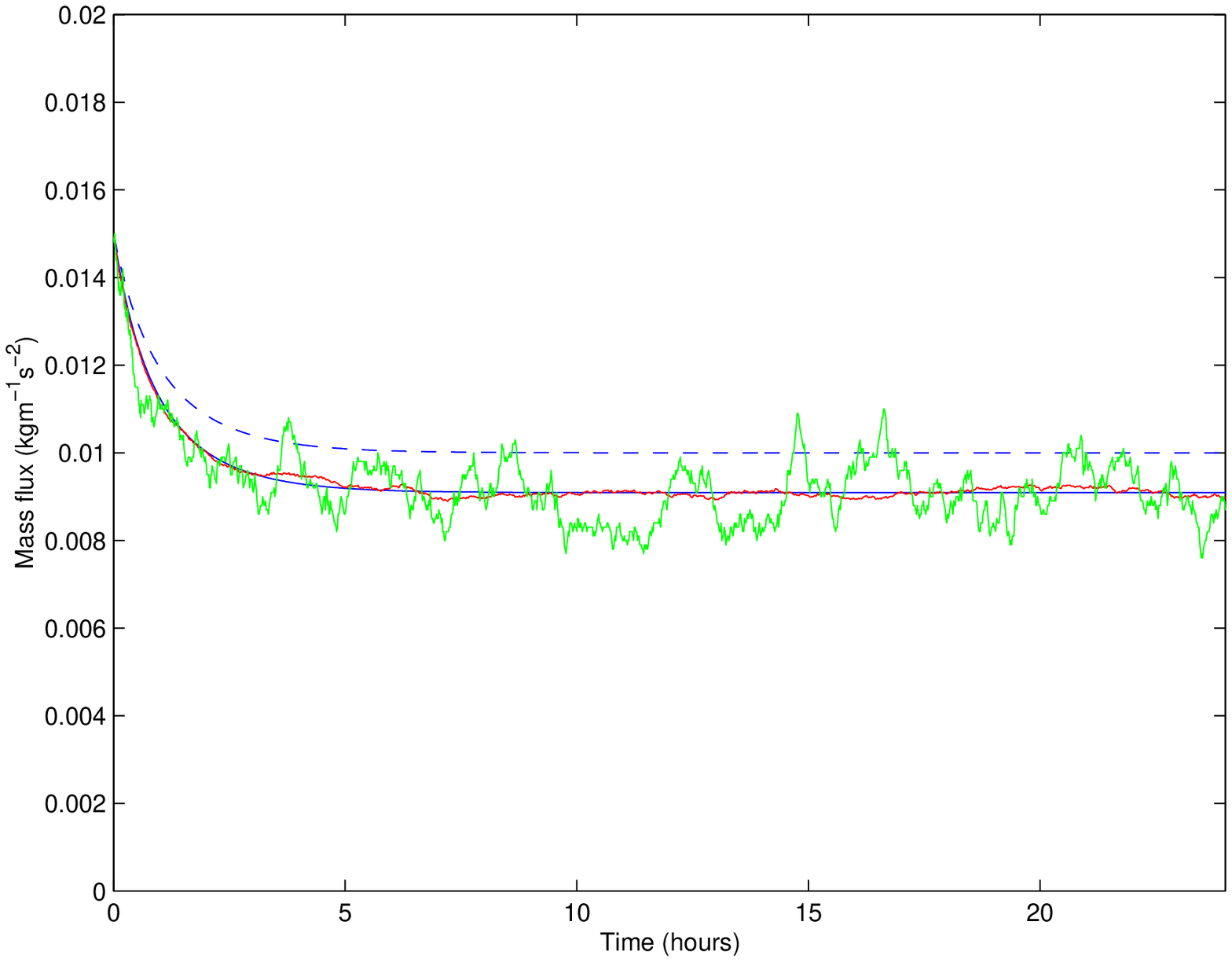}
\includegraphics[width=0.3\textwidth,height=0.25\textheight]{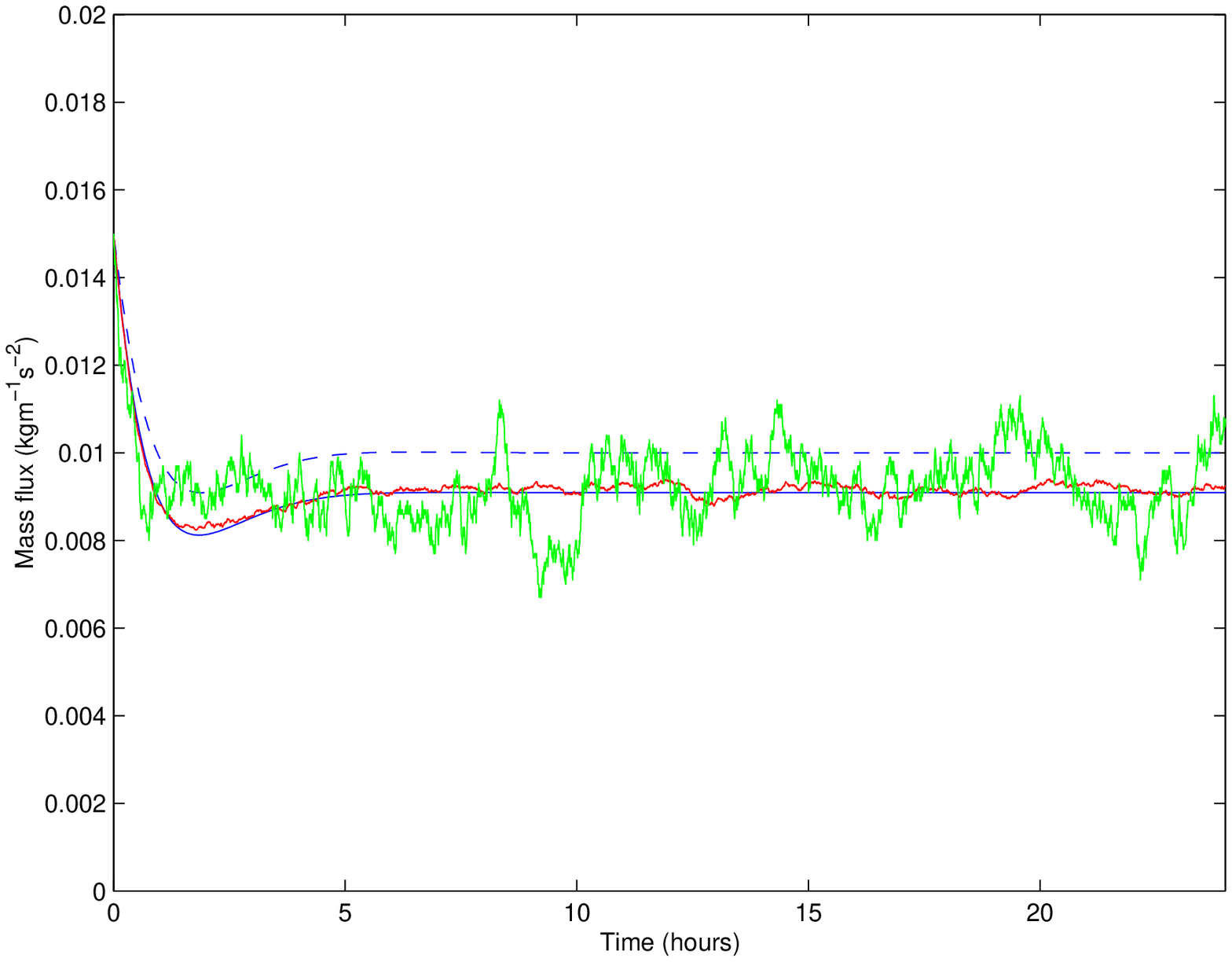}
\includegraphics[width=0.3\textwidth,height=0.25\textheight]{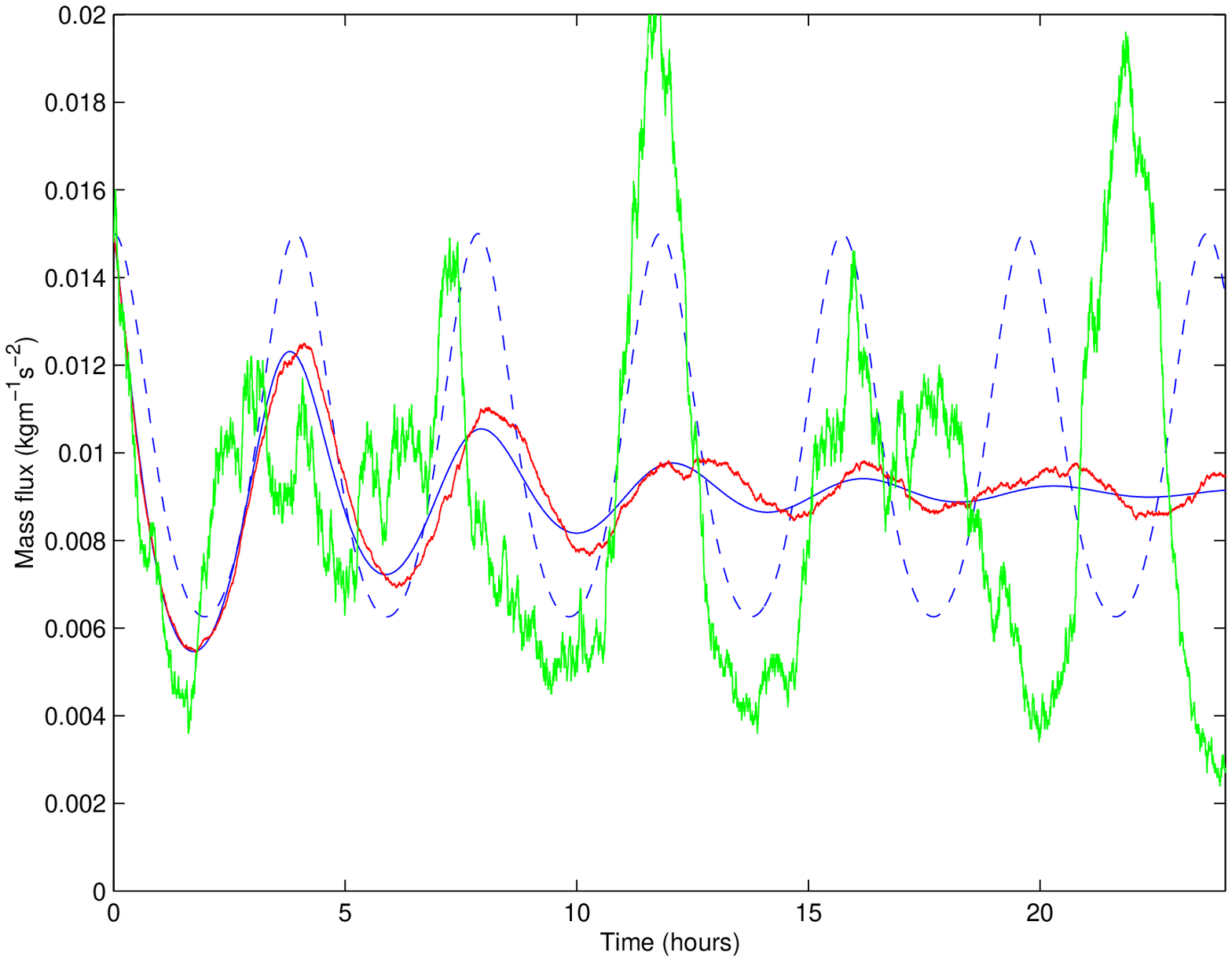}
\caption{Time series of cloud-base mass-flux for the population dynamics system \cite{popdyn} (left), the Pan and Randall system \cite{panrandall,randall+pan_mono} (centre) and the Yano and Plant system \cite{p=1} (right). The dashed blue line is the solution of the appropriate ordinary differential equation with initial condition $A=50$\,J\,kg$^{-1}$, $M_B=0.015$\,kg\,m$^{-2}$\,s$^{-1}$ and parameters as stated in the main text. The solid blue line is the solution to a slightly-modified ordinary differential equation with small cloud fractional area no longer assumed. The solid green line is a single example realization of the minimal equivalent individual-level model. The solid red line is the ensemble mean from 100 realizations of the individual-level model.\label{tseries}}
\end{figure}

Fig.~\ref{tseries} shows time series of cloud-base mass-flux from the three systems, including both individual-level results and the results from the macroscopic equations. The individual-level models do not reproduce the equilibrium state predicted by the macroscopic systems of Sec.~\ref{picture}\ref{prog} since those macroscopic systems assume a vanishing cloud fractional area. However, it is straightforward to modify those systems to account for finite cloud fractional area. One can simply take the minimal individual-level model necessary to produce the appropriate form of Eqs.~\ref{sigma_lim} and \ref{A_lim} and then apply the choices of processes and parameter settings to the complete macroscopic ordinary differential equations, as given by Eqs.~\ref{sigma_full} and \ref{A_full}. The results from the ensemble-mean of the individual-level models agree very well with these modified macroscopic systems, as indeed they should for a large enough system. This is despite the fact that there are very clear fluctuations in the timeseries from individual realizations. The difference between the prognostic system of Sec.~\ref{picture}\ref{prog} and a realization of the individual-based model is particularly apparent for the Yano and Plant system \cite{p=1}. As noted in Sec.~\ref{picture}\ref{prog} this system exhibits a periodic cycle of convective recharge and discharge but we find here that it can be slowly driven towards equilibrium through the effects of finite cloud fractional area. Nonetheless the periodic cycle remains mainfest even in longer simulations with initial transients removed, a power spectrum of the fluctuations showing a peak associated with the orbital period of $2\pi \sqrt{\beta/F}$ \cite{p=1} which is $4$hr for the parameters used here.

\begin{figure}[htb]
\centering
\includegraphics[width=0.3\textwidth,height=0.25\textheight]{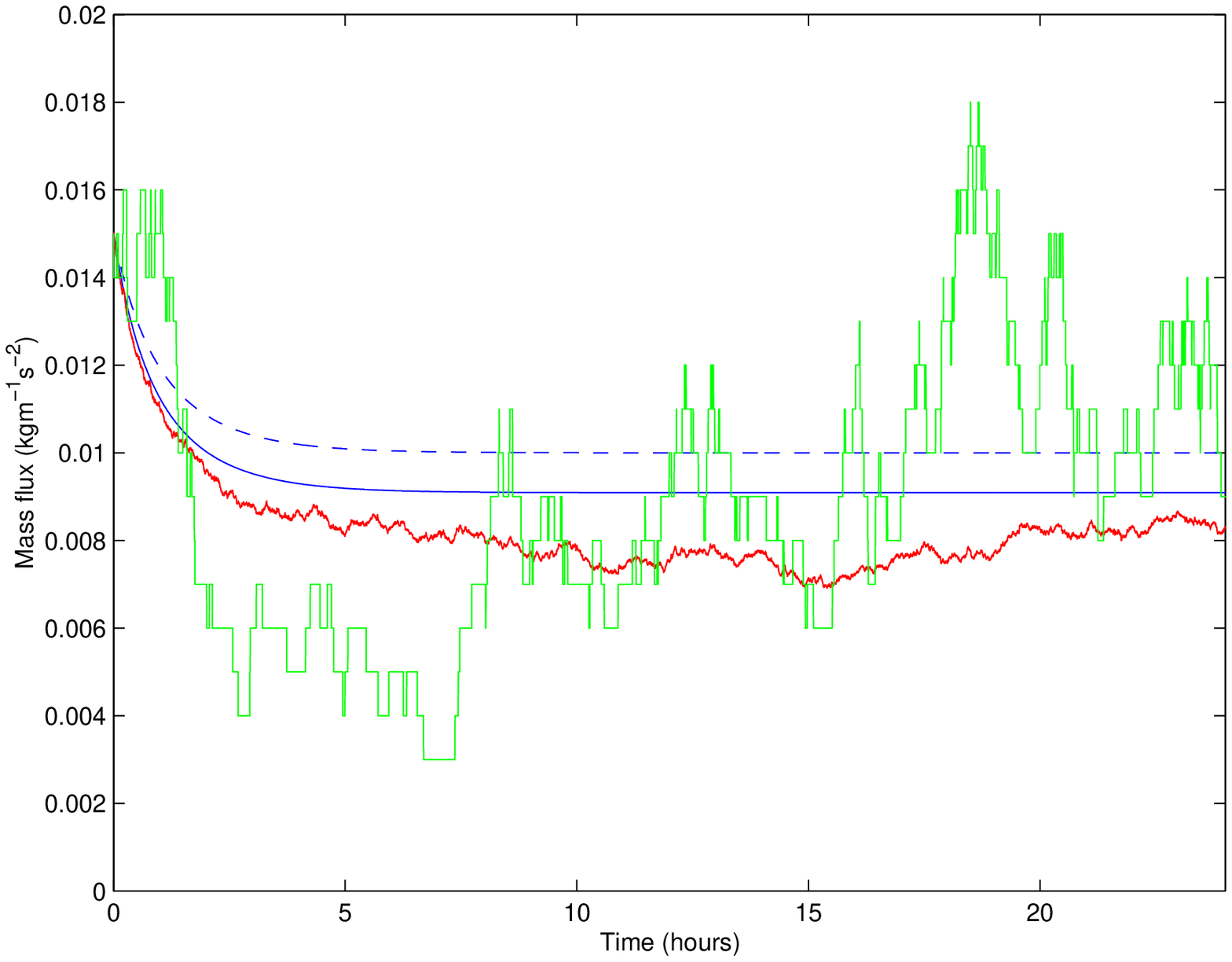}
\includegraphics[width=0.3\textwidth,height=0.25\textheight]{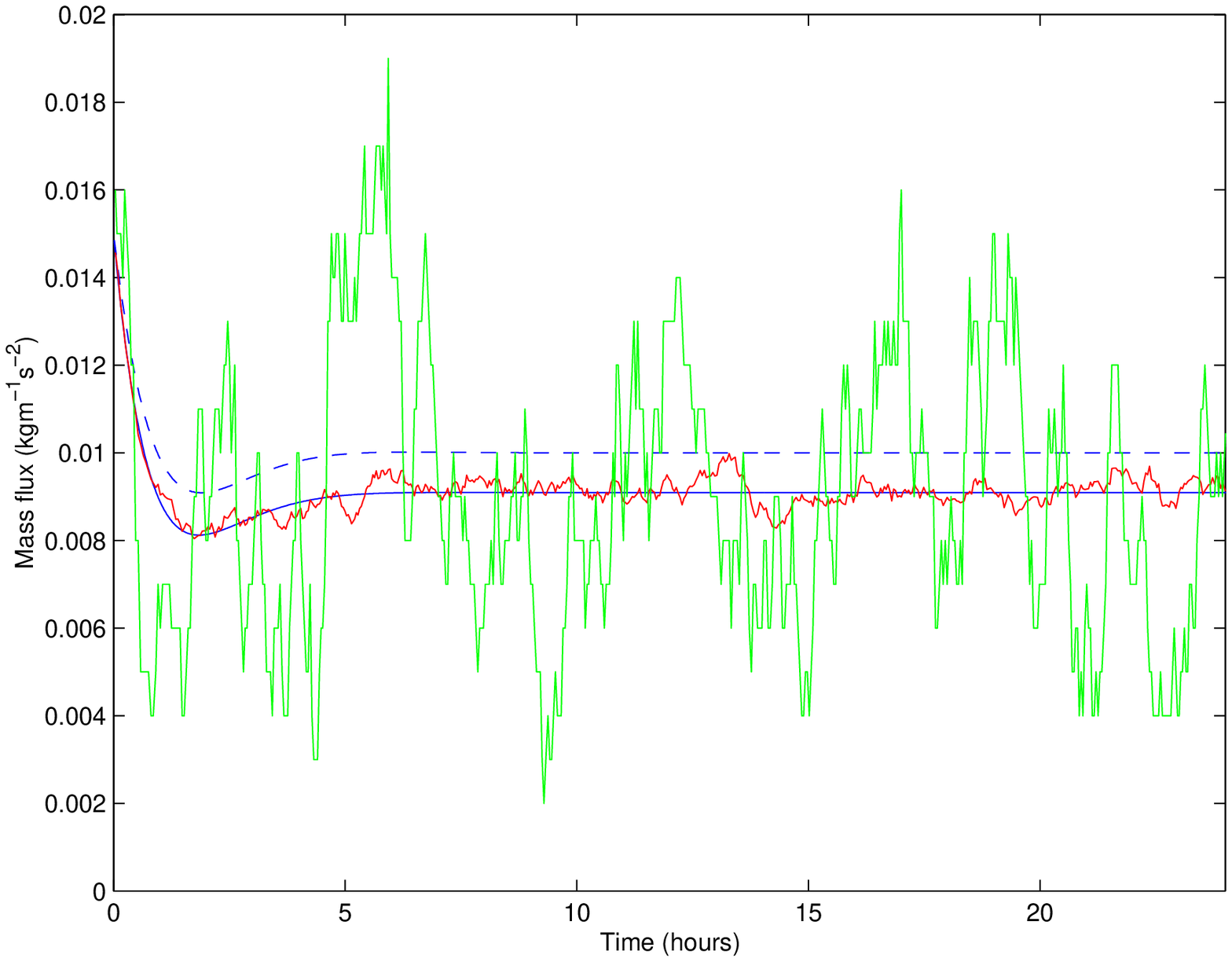}
\includegraphics[width=0.3\textwidth,height=0.25\textheight]{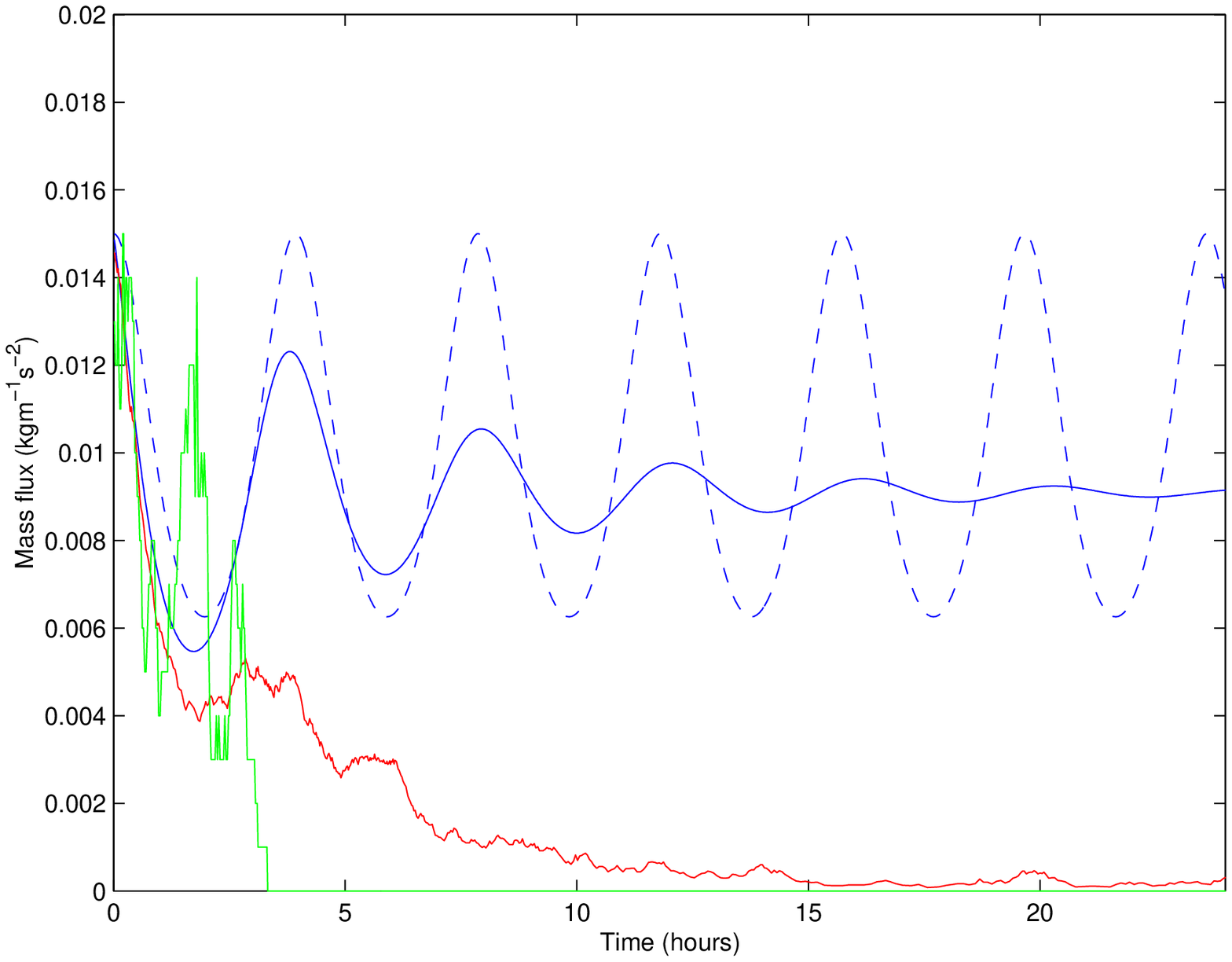}
\caption{As in Fig.~\ref{tseries} but for a smaller domain of size $\Omega=100$.\label{tseries_small}}
\end{figure}

Results for a smaller system size of $\Omega=100$ are shown in Fig.~\ref{tseries_small}. For this domain the individual-level model for the population dynamics system does show some departures from the macroscopic limit, which is however more closely respected by the equivalent to the Pan and Randall system. For the individual-level equivalent of the Yano and Plant system, convective activity dies off completely after a few hours of the example realization, never to be resumed. Convection was extinguished in all 100 realizations by $28$\,hr of simulation. For this domain size and with these parameter settings, the fluctuations in the Yano and Plant equivalent are strong enough to be able occasionally to remove all clouds present. This microscopic model does not permit the convective cloud field to recover from such an eventuality since cloud formation may only occur if induced by pre-existing clouds.

\begin{figure}[htb]
\centering
\includegraphics[width=0.3\textwidth,height=0.25\textheight]{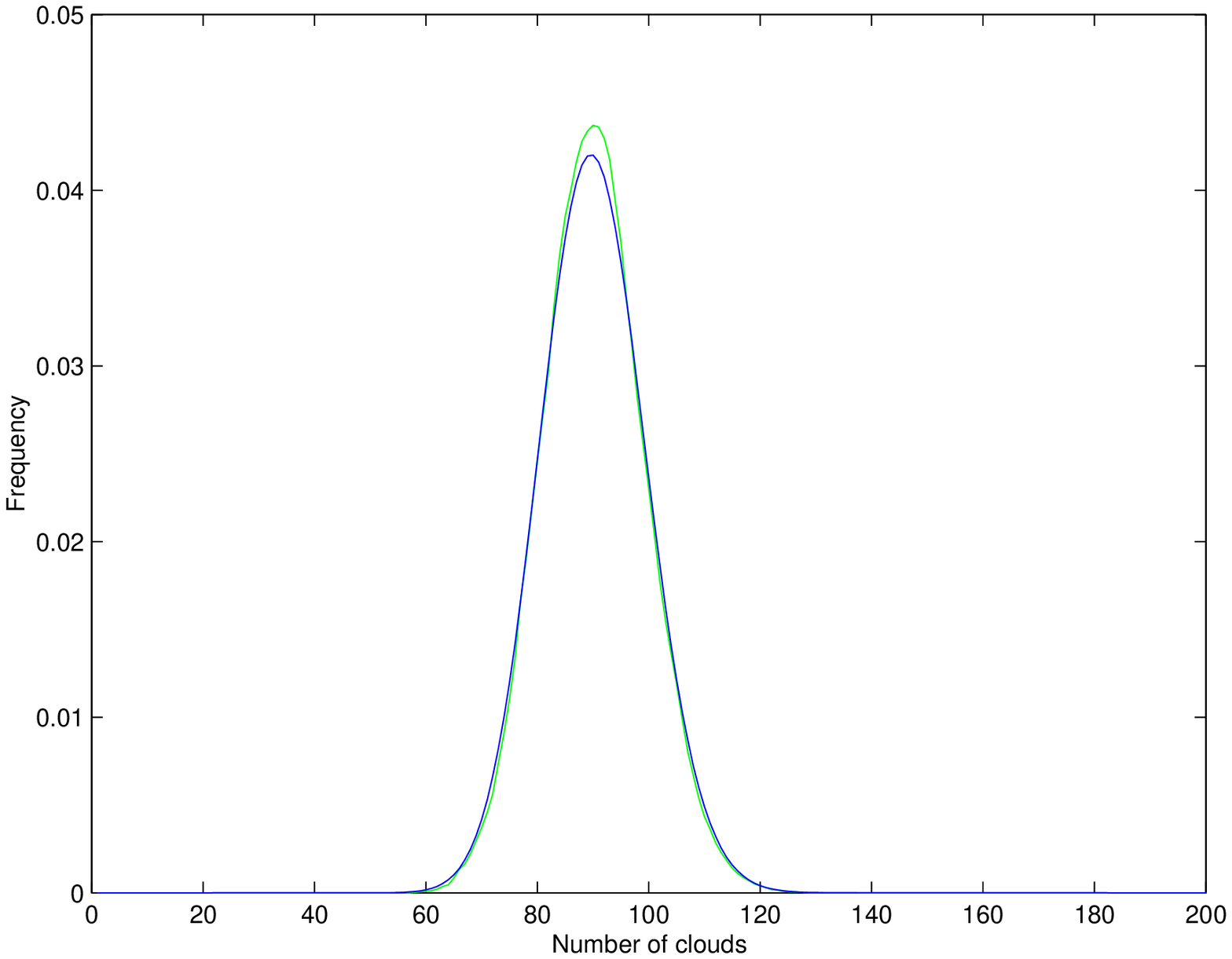}
\includegraphics[width=0.3\textwidth,height=0.25\textheight]{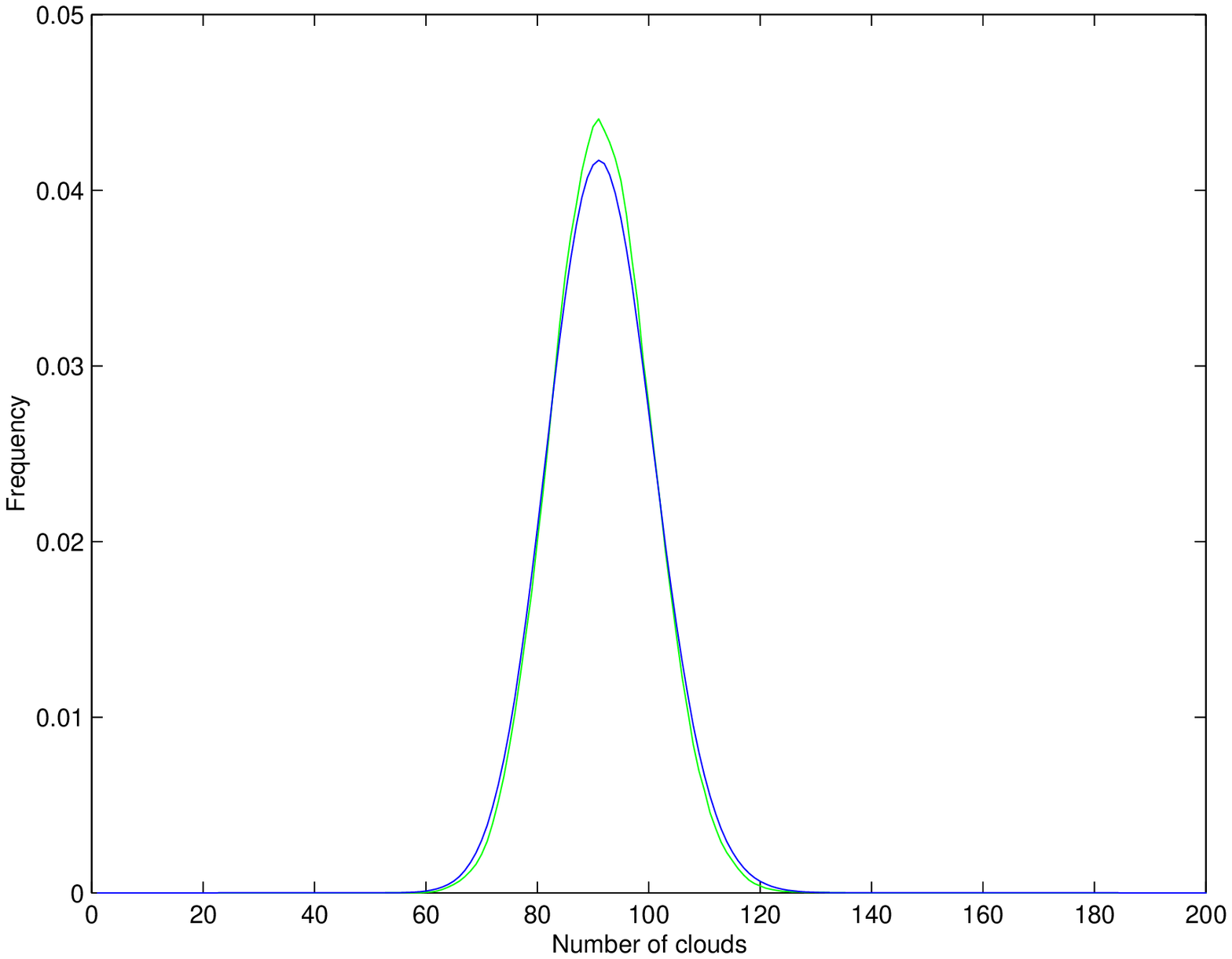}
\includegraphics[width=0.3\textwidth,height=0.25\textheight]{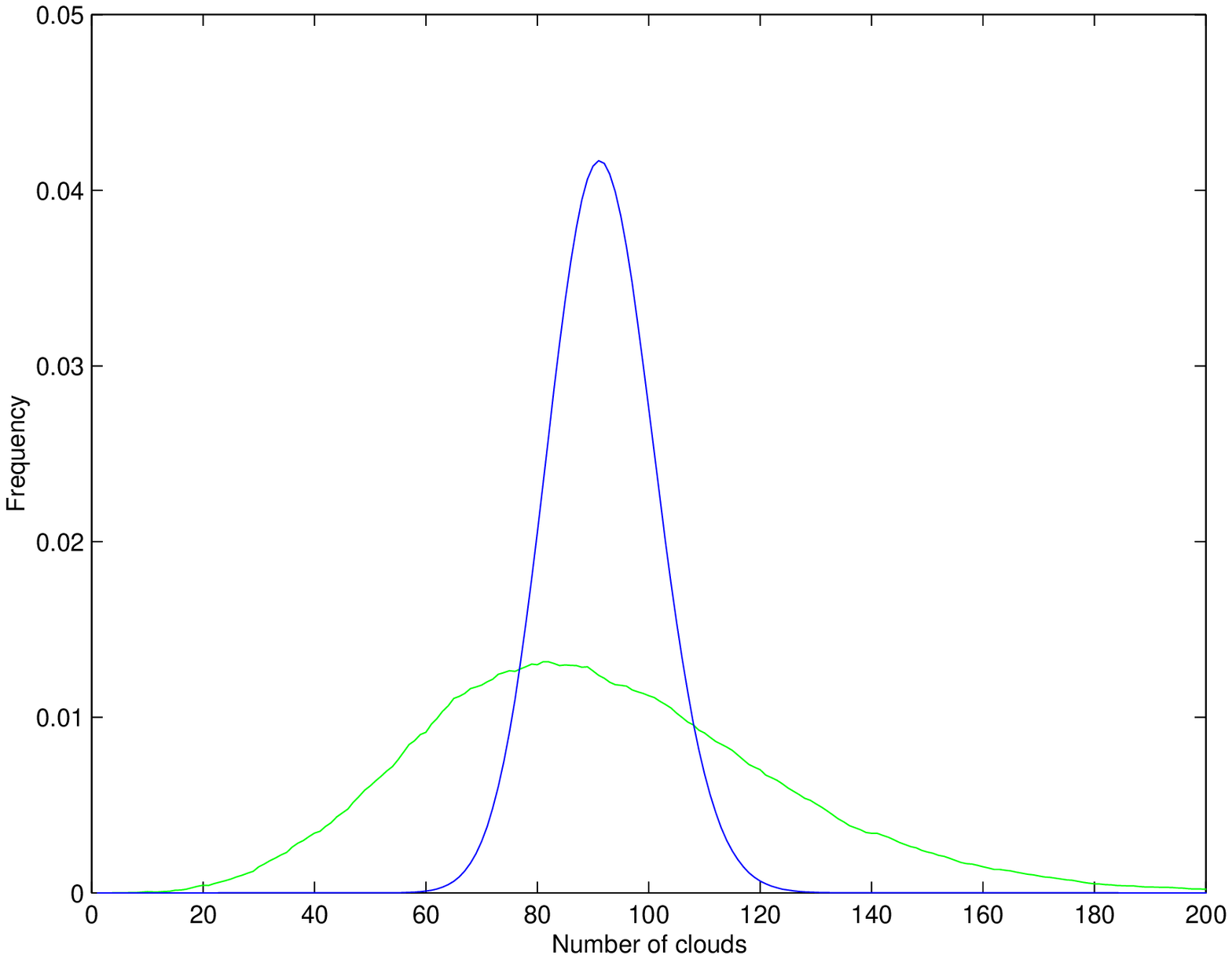}
\caption{Probability distribution functions for the  number of clouds in the \change{equilibrium state}{minimal equivalent microscopic mdels}. Results are shown for the population dynamics system \cite{popdyn} (left), the Pan and Randall system \cite{panrandall,randall+pan_mono} (centre) and the Yano and Plant system \cite{p=1} (right). The green line is constructed from a single example realization of the minimal equivalent individual-level model, using data between $48$ and $120$ hours of simulation. The blue line is a Poisson distribution for the same mean cloud number.\label{pdf_n}}
\end{figure}

Returning now to the $\Omega=1000$ domain, probability distribution functions for the number of clouds present \remove{in the equilibrium state} are shown in Fig.~\ref{pdf_n}. For the population dynamics \cite{popdyn} and Pan and Randall systems \cite{panrandall,randall+pan_mono}, the results are very well approximated by a Poisson distribution, in accordance with the theoretical expectations of Craig and Cohen \cite{craigcohen} (Sec.~\ref{picture}\ref{stoch}). In contrast, the distribution from the individual-level equivalent of the Yano and Plant system \cite{p=1} is much wider. This system was not designed to produce a highly-stable equilibrium state but rather to demonstrate the cycles of recharge and discharge that are characteristic of some convective systems. The different distributions for cloud number can be understood in terms of the different mechanisms for cloud formation in the equivalent individual-level models. For the equivalent to the \remove{population dynamics and} Pan and Randall system clouds are formed spontaneously at empty \change{sites on the lattice}{elements}, and the number of such empty \change{sites}{elements} deviates only weakly from $E=\Omega-N \approx \Omega$. By contrast, in the equivalent to the \add{population dynamics and} Yano and Plant systems, clouds are formed only in association with preexisting clouds: \change{as we have seen, the number of these}{$N$ is much more susceptible to fluctuations, and the formation mechanism will itself tend to amplify the fluctuations. The population dynamics system, however, has a compensating mechanism because that the removal rate from competitive exclusion also depends on the number of preexisting clouds.}

In reality, both primary and secondary mechanisms of cumulus cloud formation do occur, the extent of each being rather sensitive to the prevailing meteorological and topographical conditions \cite{bennett}. \add{One might therefore reasonably expect that each of the macroscopic models investigated here could perform well in different limiting cases, where different microscopic processes are more or less important. To capture the full range of convective behaviours, a hybrid of the existing macroscopic models would presumably be needed.}

\section{Conclusions}\label{concl}

The description of atmospheric convection as a collection of distinct plumes has a long history. It is an instructive basis from which to seek to understand many features of convective systems, and still forms the underlying basis for most current convective parameterizations. Some brutal simplifying assumptions have usually been imposed in such studies, but there has been an increasing recognition in recent years that some of those simplifications may be neither necessary nor desirable. Satisfactory models for statistical cumulus dynamics remain to be developed, and tools from statistical physics are likely to be required to do so.

In this article we have proposed a new modelling framework well-suited to the description of collections of convective plumes. In doing so we have been mindful of two common simplifications in particular: the assumptions of convective quasi-equilibrium and of large cloud numbers. However, we believe the framework to be easily extendable \cite{mckane04,lugo} to examine other important issues in atmospheric convection: for example, the role of various (self-)organizational mechanisms in developing spatial structure. Previous authors have developed models for the time evolution of convection and for stochastic effects \change{due to finite cloud numbers, but the present work is the first attempt to marry those earlier models in a unified description that is both stochastic and prognostic from the outset.}{\protect\cite{randall+pan_mono,kmk03,mk02,kbm10,craigcohen,randallhuffman,panrandall,popdyn,p=1}. The present work attempts to marry some of those earlier models in a unified description that is both stochastic and prognostic from the outset.}

The modelling framework is developed from the individual level of single clouds. Each cloud is treated identically and extremely simply here, and is formed, modifies its environment and meets its demise according to straightfoward probabilistic rules. Doubtless there is much scope for elaboration on each of these points. Thus, we prefer to speak of a framework rather than a complete model for SCD. However, the simplest treatment is quite sufficient \change{to make contact with the previous studies in the area}{for the present aims}. Great stress has been placed throughout on the notion that the individual-level model should reduce to the systems of \add{some} previous studies in the appropriate limits. By means of van Kampen's system-size expansion we can show that by making appropriate choices of the processes included in the individual-level model, we can recover previous prognostic models in the limits of a large system size and a vanishing cloud fractional area. Moreover, by making appropriate choices of the processes that form clouds, we can also recover previous predictions for fluctuations in cloud number at equilibrium.

As a result, we assert that the proposed framework has been well established as a methodology that both encompasses and extends current attempts to develop a theory of SCD. For instance, we have already been able to gain some insights into previously-proposed prognostic systems by establishing and simulating their equivalent individual-based models. It is certainly not obvious from the original articles that the primary difference between the Pan and Randall \cite{panrandall,randall+pan_mono} and Yano and Plant \cite{p=1} systems is an implicit assumption about the dominant microscopic \change{mechanism for}{process of} convective cloud initiation. Intermediate models which admit both \change{mechanisms}{processes} would seem more physically reasonable, could very easily be built in the present framework, and would be well worthy of further investigation.

\begin{acknowledgements} 
Useful discussions with Jun-Ichi Yano are acknowledged, and were enabled by a joint project award by the Royal Society and CNRS. Some of the work was conducted during a visiting fellowship at the Isaac Newton Institute for Mathematical Sciences as part of their programme on Mathematical and Statistical Approaches to Climate Modelling and Prediction. I am grateful for many discussions there, for various discussions supported through COST Action ES0905, and also for constructive suggestions on the manuscript from the referees.
\end{acknowledgements}

\setlength{\bibsep}{0pt}
\bibliographystyle{/home/sws00rsp/latex/royalsoc/rspublicnat}
\bibliography{revised}

\label{lastpage}
\end{document}